\documentstyle[twocolumn,epsf]{mn}

\def\spose#1{\hbox to 0pt{#1\hss}}
\def\lta{\mathrel{\spose{\lower 3pt\hbox{$\sim$}}
    \raise 2.0pt\hbox{$<$}}}
\def\gta{\mathrel{\spose{\lower 3pt\hbox{$\sim$}}
    \raise 2.0pt\hbox{$>$}}}
\def\farcm{\hbox{$.\mkern-4mu^\prime$}}
\def\farcs{\hbox{$.\!\!^{\prime\prime}$}}
\def\kms{$\mbox{km~s}^{-1}$}
\def\micron{$\mu\mbox{m}$}
\def\fig#1{\centerline{\epsffile{#1}}}
\def\sauron{\texttt{SAURON}}
\def\oasis{\texttt{OASIS}}
\def\tiger{\texttt{TIGER}}
\def\vimos{\texttt{VIMOS}}
\def\integral{\texttt{INTEGRAL}}

\def\stis{\texttt{STIS}}

\def\wfpc2{\texttt{WFPC2}}
\def\etal{\emph{et~al.}}
\title[The SAURON project]
{The SAURON project. I. The panoramic integral-field spectrograph}
\author[R. Bacon \etal]
{R. Bacon$^1$\thanks{E-mail: bacon@obs.univ-lyon1.fr},
  Y.\ Copin$^{1, 2}$, G.\ Monnet$^3$, Bryan W.\ Miller$^4$, 
  J.R.\ Allington-Smith$^5$, 
  \newauthor
  M.\ Bureau$^2$, C.\ Marcella Carollo$^6$, Roger L.\ Davies$^5$, 
  Eric Emsellem$^1$, 
  \newauthor
  Harald Kuntschner$^5$, Reynier F.\ Peletier$^7$, 
  E.K.\ Verolme$^2$, P.\ Tim de Zeeuw$^2$ \\
$1$ CRAL-Observatoire, 9 Avenue Charles Andr\'e, 
    69230 Saint-Genis-Laval, France\\
$2$ Sterrewacht Leiden, Niels Bohrweg 2, 2333 CA, Leiden, The Netherlands \\
$3$ European Southern Observatory, Karl-Schwarzschild Strasse 2, 
    D-85748 Garching, Germany\\
$4$ Gemini Observatory, Casilla 603, La Serena, Chile \\
$5$ Physics Departement, University of Durham, 
    South Road, Durham DH13LE, United Kingdom \\
$6$ Department of Astronomy, Columbia University, 
    538 West 120th Street, New York, NY 10027, USA \\
$7$ Department of Physics and Astronomy, University of Nottingham, 
    University Park, Nottingham NG7 2RD, United Kingdom
}

\date{Accepted $\ldots$ 
      Received $\ldots$;
      in original form $\ldots$} 

\begin{document}
\maketitle

\begin{abstract}
A new integral-field spectrograph, \sauron, is described. It is based
on the \tiger\ principle, and uses a lenslet array.  \sauron\ has a
large field of view and high throughput, and allows simultaneous sky
subtraction. Its design is optimized for studies of the stellar
kinematics, gas kinematics, and line-strength distributions of nearby
early-type galaxies. The instrument design and specifications are
described, as well as the extensive analysis software which was
developed to obtain fully calibrated spectra, and the associated
kinematic and line-strength measurements. A companion paper reports on
the first results obtained with \sauron\ on the William Herschel
Telescope.
\end{abstract}


\section{Introduction}
\label{sec:intro}
Determining the intrinsic shapes and internal dynamical and stellar
population structure of elliptical galaxies and spiral bulges is a
long-standing problem, whose solution is tied to understanding the
processes of galaxy formation and evolution (Franx, Illingworth \& de
Zeeuw 1991; Bak \& Statler 2000).  Many galaxy components are not
spherical or even axisymmetric, but triaxial: N-body simulations
routinely produce triaxial dark halos (Barnes 1994; Weil \& Hernquist
1996); giant ellipticals are known to be slowly-rotating triaxial
structures (Binney 1976, 1978; Davies \etal\ 1983; Bender \& Nieto
1990; de Zeeuw \& Franx 1991); many bulges, including the one in the
Galaxy, are triaxial (Stark 1977; Gerhard, Vietri \& Kent 1989;
H\"afner \etal\ 2000); bars are strongly triaxial (Kent 1990;
Merrifield \& Kuijken 1995; Bureau \& Freeman 1999). Key questions for
scenarios of galaxy formation include: What is the distribution of
intrinsic triaxial shapes?  What is, at a given shape, the range in
internal velocity distributions?  What is the relation between the
kinematics of stars (and gas) and the stellar populations?  Answers to
these questions require morphological, kinematical and stellar
population studies of galaxies along the Hubble sequence, combined
with comprehensive dynamical modeling.\looseness=-2

The velocity fields and line-strength distributions of triaxial galaxy
components can display a rich structure, that are difficult to map
with traditional long-slit spectroscopy (Arnold \etal\ 1994; Statler
1991, 1994; de Zeeuw 1996). Furthermore, in many elliptical galaxies,
the central region is `kinematically decoupled': the inner and outer
regions appear to rotate around different axes (e.g., Franx \&
Illingworth 1988; Bender 1988; de Zeeuw \& Franx 1991; Surma \& Bender
1995). This makes \emph{two-dimensional} (integral-field) spectroscopy
of stars and gas essential for deriving the dynamical structure of
these systems and for understanding their formation and evolution.

\begin{figure*}
\epsfxsize=\textwidth
\fig{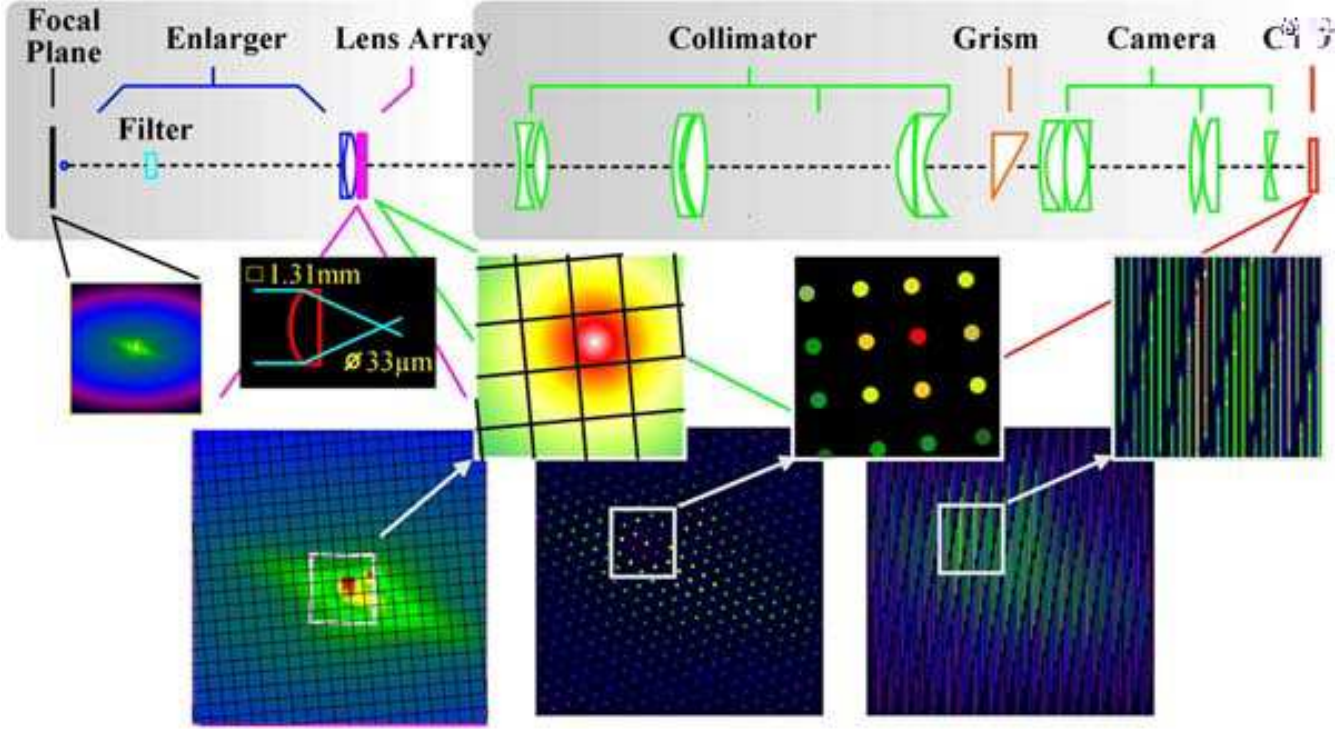}
\caption{Optical layout of \sauron. The main optical elements are
displayed from the telescope focal plane (left) to the detector plane
(right).  The image of a galaxy is shown successively at the telescope
focal plane, at the entrance plane of the lens array, at the exit
plane of the lens array and at the detector plane. A zoom-in of a
small area of the galaxy is shown at each plane. The insert shows the
detailed light path within a microlens. }
\label{fig:layout}
\end{figure*}

In the past decade, substantial instrumental effort has gone into
high-spatial resolution spectroscopy with slits (e.g., \stis\ on HST,
Woodgate \etal\ 1998), or with integral-field spectrographs with small
field of view (e.g., \tiger\ and \oasis\ on the CFHT; Bacon \etal\
1995, 2000) to study galactic nuclei. By contrast, studies of the
large-scale kinematics of galaxies still make do with long-slit
spectroscopy along at most a few position angles (e.g., Davies \&
Birkinshaw 1988; Statler \& Smecker-Hane 1999).  While useful for
probing the dark matter content at large radii (e.g., Carollo \etal\
1995; Gerhard \etal\ 1998), long-slit spectroscopy provides
insufficient spatial coverage to unravel the kinematics and
line-strength distributions of early-type galaxies. Therefore, we
decided to build \sauron\ (Spectroscopic Areal Unit for Research on
Optical Nebulae), an integral-field spectrograph optimized for studies
of the kinematics of gas and stars in galaxies, with high throughput
and, most importantly, with a large field of view.  We are using
\sauron\ on the 4.2m William Herschel Telescope (WHT) on La Palma to
measure the mean streaming velocity $V$, the velocity dispersion
$\sigma$ and the velocity profile (or line-of-sight velocity
distribution, LOSVD) of the stellar absorption lines and the emission
lines of ionized gas, as well as the two-dimensional distribution of
line-strengths and line-ratios, over the optical extent of a
representative sample of nearby early-type galaxies.  The first
results of this project are described in Davies \etal\ (2001) and de
Zeeuw \etal\ (2001, Paper~II).\looseness=-2

This paper presents \sauron, and is organized as follows:
\S\ref{sec:instrument} describes the design, the choice of the
instrumental parameters and the construction of the instrument; the
results of commissioning and instrument calibration are reported in
\S\ref{sec:commissioning}, and the extensive set of data reduction
software we developed to calibrate and analyze \sauron\ exposures are
detailed in \S\ref{sec:datared} and \S\ref{sec:dataana};
\S\ref{sec:example} presents a first result, and
\S\ref{sec:conclusions} sums up.\looseness=-2

\section{The SAURON spectrograph}
\label{sec:instrument}

The design of \sauron\ is similar to that of the prototype
integral-field spectrograph \tiger, which operated on the CFHT between
1987 and 1996 (Bacon \etal\ 1995), and of \oasis, a common-user
integral-field spectrograph for use with natural guide star adaptive
optics at the CFHT (Bacon \etal\ 2000). Fig.~\ref{fig:layout}
illustrates the optical layout: a filter selects a fixed wavelength
range, after which an enlarger images the sky onto the heart of the
instrument, a lenslet array.  Each lenslet produces a micropupil,
whose light, after a collimator, is dispersed by a grism.  A camera
then images the resulting spectra onto a CCD, aligned with the
dispersion direction but slightly rotated with
respect to the lenslet array to avoid overlap of adjacent spectra.

\subsection{Setting instrumental parameters}
\label{sec:setting}

Compared to \oasis, which is dedicated to high spatial resolution, a
large field of view is the main driver of the \sauron\ design.  There
are a number of free parameters which can be used to maximize the
field of view: the spectral length in CCD pixels, the perpendicular
spacing between adjacent spectra (which we refer to as the
cross-dispersion separation), the detector size and the lenslet size
on the sky. The latter is related to the diameter of the telescope and
the aperture of the camera for a given detector pixel size (see
eq.~[12] of Bacon \etal\ 1995). For the 4.2m aperture of the WHT and
the 13.5~\micron\ pixels of the 2k$\times$4k EEV~12 detector this
limits the lenslet size to a maximum of about $1''$ in order to keep
the camera aperture slower than $f/1.8$.

\begin{table}
\caption{Current specification of \sauron\ on the WHT.}
\label{tab:spec}
\begin{tabular}{lcc}
\hline\hline
                           & HR mode           & LR mode \\
\hline
Spatial sampling           & $0\farcs27$       & $0\farcs94$ \\
Field of view              & $9'' \times 11''$ & $33'' \times 41''$  \\
Spectral resolution (FWHM) & 2.8~\AA           & 3.6~\AA \\
Instrumental dispersion ($\sigma_{\mathrm{inst}}$) & 90~\kms   & 105~\kms \\
\noalign{\medskip}
Spectral sampling          & \multicolumn{2}{|c|}{1.1~\AA\ pix$^{-1}$} \\
Wavelength range           & \multicolumn{2}{|c|}{4500--7000~\AA} \\
Initial spectral window    & \multicolumn{2}{|c|}{4760--5400~\AA} \\
Calibration lamps          & \multicolumn{2}{|c|}{Ne, Ar, W} \\
Detector                   & \multicolumn{2}{|c|}{EEV~12 $2148\times 4200$} \\
Pixel size                 & \multicolumn{2}{|c|}{13.5~\micron} \\
Instrument Efficiency      & \multicolumn{2}{|c|}{35\%} \\
Total Efficiency           & \multicolumn{2}{|c|}{14.7\%} \\
\hline
\end{tabular}
\end{table}

The cross-dispersion separation is a critical parameter. Its value is
the result of a compromise: on the one hand, it should be as small as
possible in order to maximize the number of spectra on the detector;
on the other hand, it should be large enough in order to avoid overlap
of neighbouring spectra.  For \sauron, the chosen value of 4.8~pixels
gives up to 10\% overlap between adjacent spectra (see
Fig.~\ref{fig:crossdisp}).
The trade-off for this small cross-dispersion separation was to
invest significant effort in the development of specific software to
account for this overlap at the stage of spectra extraction (see
\S\ref{sec:extraction}).  Given the fast aperture of the camera, we
decided to limit its field size by using only 2k$\times$3k pixels on
the detector. The required image quality can then be obtained with a
reasonable number of surfaces, so that the throughput is as high as
possible.\looseness=-2

The spectral length is the product of the spectral coverage and the
spectral sampling. We selected the 4800--5400~\AA\ range so as to
allow simultaneous observation of the [O\textsc{iii}] and H$\beta$
emission lines to probe the gas kinematics, and of a number of
absorption features (the Mg$b$ band, various Fe lines and again
H$\beta$) for measurement of the stellar kinematics (mean velocity,
velocity dispersion and the full LOSVD) and
line-strengths.\footnote{In the future, we plan to use \sauron\ in
other spectral windows in the range 4500--7000~\AA\ (see
\S\ref{sec:design}), including, e.g., the H$\alpha$ line.} The adopted
spectral dispersion of \sauron\ in the standard mode (see below) is
$\sigma_{\mathrm{inst}} \simeq 90$~\kms (with a corresponding sampling
of 1.1~\AA~pixel$^{-1}$), which is adequate to measure the LOSVD of
the stellar spheroidal components of galaxies which have velocity
dispersions $\gta 100$~\kms.

The final result is a field of view of $41''\times33''$, observed with
a filling factor of 100\%. Each lenslet has a square shape and
samples a $0\farcs94\times0\farcs94$ area on the sky in this
low-resolution (LR) mode (Table~\ref{tab:spec}).  This undersamples
the typical seeing at La Palma, but provides essentially all one can
extract from the integrated light of a galaxy over a large field of
view in a single pointing. 

To take advantage of the best seeing conditions, \sauron\ contains a
mechanism for switching to another enlarger, resulting in a
$9''\times11''$ field of view sampled at $0\farcs27\times0\farcs27$
(Table~\ref{tab:spec}).  This high-resolution (HR) mode can be used in
conditions of excellent seeing to study galactic nuclei with the
highest spatial resolution achievable at La Palma without using
adaptive optics.

\begin{figure}
\epsfxsize=\columnwidth
\fig{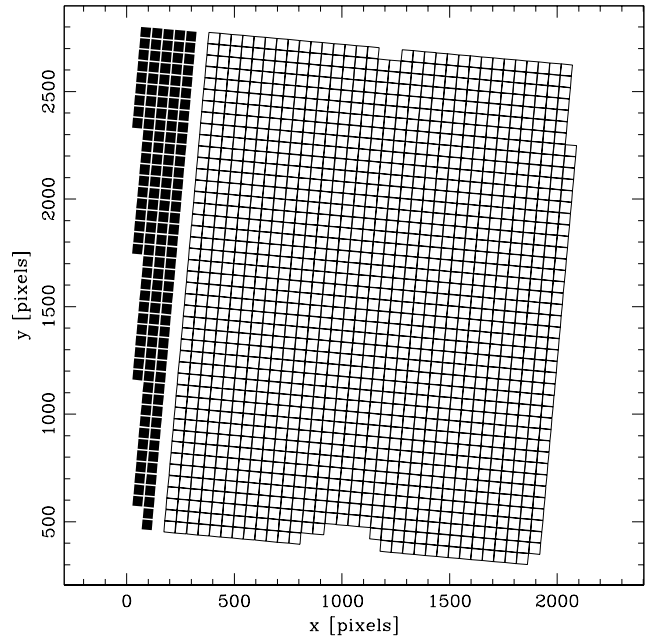}
\caption{The \sauron\ lenslet position on the CCD. Open
squares: object lenses, solid squares: sky lenses. The scale
is in pixels. The sky lenses have the same shape and size as the
object lenses (i.e., 0\farcs94 in the LR mode and 0\farcs27 in the HR
mode), but are shifted 1\farcm9 away from the center of the field.}
\label{fig:lensmap}
\end{figure}

As the objective is to map the kinematics and line-strengths out to
about an effective radius for nearby early-type galaxies, it is
necessary to have accurate sky subtraction.  However, for most
objects, the field of view is still not large enough to measure a
clean sky spectrum on it. We therefore incorporated in \sauron\ the
ability to acquire sky spectra simultaneously with the object spectra.
This is achieved by means of an extra enlarger pointing to a patch of
sky located $1\farcm9$ away from the main field: the main enlarger
images the science field of view on the center of the lens array,
while the sky field of view is imaged off to one side; a combination
of a mask and a baffle is used to prevent light pollution between the
two fields. This gives 1431 object lenslets and 146 sky lenslets
(Fig.~\ref{fig:lensmap}).

\subsection{Design and realization}
\label{sec:design}

The \sauron\ optics were designed by Bernard Delabre (ESO). The main
difficulty was to accommodate the required fast aperture of the camera
($f/1.8$).  To maximize the throughput and stay within a limited
budget, the optics were optimized for the range 4500--7000~\AA, rather
than for the whole CCD bandwidth (0.35--1~\micron).  This provides the
required image quality (80\% encircled energy within a 13~\micron\
radius) with a minimum number of surfaces and no aspheres. The
required spectroscopic resolution is obtained with a
514~lines~mm$^{-1}$ grism and a 190~mm focal-length camera.\looseness=-2

\begin{figure}
\epsfxsize=\columnwidth
\fig{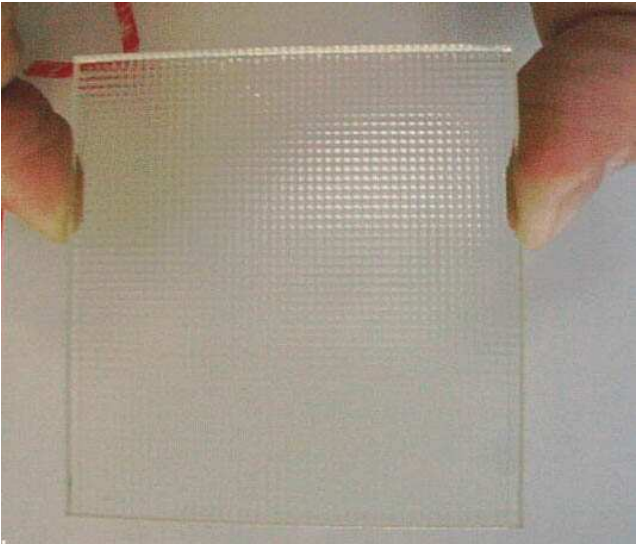}
\caption{The \sauron\ lenslet array is made from two sets of
cylindrical barrels of $1.31\times70$~mm$^2$ in fused silica. It
consists of over 1600 square lenslets, each of which is 1.31~mm on the
side.}
\label{fig:lenslet}
\end{figure}

The lens array (Fig.~\ref{fig:lenslet}) is constructed from two arrays
of cylindrical lenses mounted at right angles to each other.  This
assembly gives the geometrical equivalent of an array of square lenses
of 1.31~mm on the side.  Two identical enlargers---one for the object
and the other for the sky---with a magnification of 6.2 are used in
front of the lens array to produce the LR mode (0\farcs94 sampling).
A similar pair of enlargers with a magnification of 22.2 provides the
HR mode (0\farcs27 sampling).  An interference filter with a square
transmission profile selects the 4760--5400~\AA\ wavelength range. The
filter transmission is above 80\% over more than 80\% of this
wavelength range.  The filter is tilted to avoid ghost images and, as
a consequence, the transmitted bandwidth moves within the field.  As
will be discussed in \S\ref{sec:merge}, this reduces the useful common
bandwidth by 16\%.
   
\begin{figure}
\epsfxsize=\columnwidth
\fig{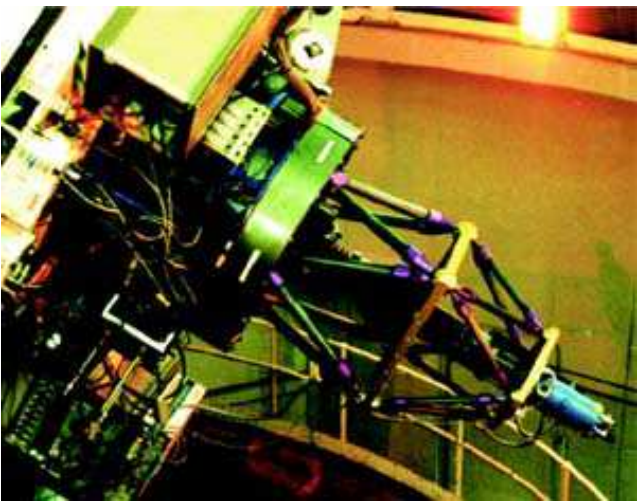}
\caption{\sauron\ at the Cassegrain port of the WHT during the
commissioning run in February 1999.}
\label{fig:instrument}
\end{figure}

The mechanical design of \sauron\ is similar to that of \oasis: a
double octapod mounted on a central plate which supports the main
optical elements.  This design is light and strong, and minimizes
flexure (cf. \S\ref{sec:commissioning}).  Fig.~\ref{fig:instrument}
shows the instrument mounted on the Cassegrain port of the
WHT.\looseness=-2

Because of its limited number of observing modes (LR/HR), \sauron\ has
only three mechanisms: the focal plane wheel, for the calibration
mirror, the enlarger wheel, which supports the two sets of twin
enlargers, and a focusing mechanism for the camera. The latter has a
linear precision and repeatibility of a few \micron, which is required
to ensure a correct focus at $f/1.8$. The temperature at the camera
location is monitored during the night. 

Calibration is performed internally using a tungsten lamp as continuum
source, and neon and argon arc lamps as reference for wavelength
calibration. Special care is taken in the calibration path to form an
equivalent of the telescope pupil with the correct figure and
location.  This is particularly relevant for \tiger-type
integral-field spectrographs, since pupils are imaged onto the
detector.

All movements are controlled remotely with dedicated electronics.
Commands are sent by a RS432 link from the control room, where a Linux
PC is used to drive the instrument. A graphical user interface (GUI)
written in Tcl/Tk allows every configuration changes to be made with
ease.

\section{Commissioning}
\label{sec:commissioning}

\sauron\ was commissioned on the WHT in early February 1999. In the
preceding days, the instrument, together with the control software,
was checked and interfaced to the WHT environment.  Particular
attention was paid to the optical alignment between \sauron\ and the
detector (dispersion direction along the CCD columns and
$5.18^{\circ}$ rotation between dispersion direction and lenslet
array). The most labour-intensive part was to adjust the CCD to be
coplanar with the camera focal plane, so that there is no detectable
systematic gradient in the measured point spread function (PSF). This
must be done accurately since a small residual tilt would rapidly
degrade the PSF within the illuminated area of the CCD. A specific procedure
was developed to ease this part of the setup.  The instrument
performed very well from the moment of first light on February 1st (de
Zeeuw \etal\ 2000).

\begin{figure}
\epsfxsize=\columnwidth
\fig{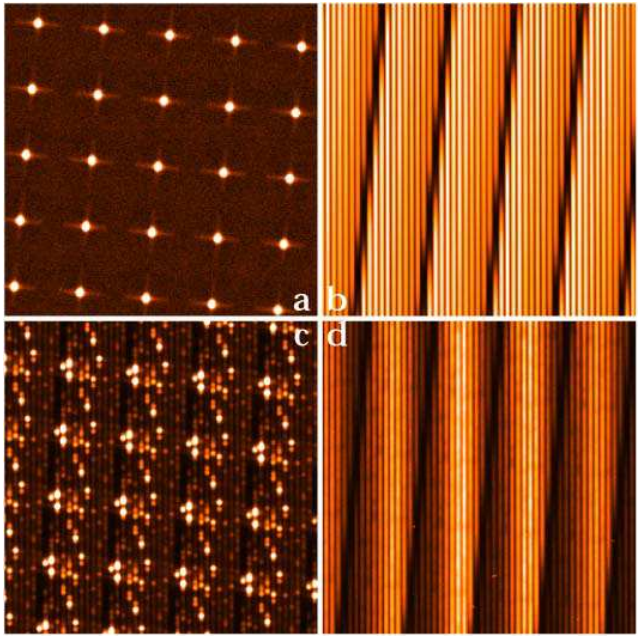}
\caption{Examples of \sauron\ exposures taken during commissioning.
Each panel shows only a small part of the entire CCD frame so that
details can be seen. a) Micropupil image taken with the grism
out. Each cross is the diffraction pattern from one square lenslet.
b) Continuum image using the tungsten lamp. c) Neon arc.  d) Central
part of NGC~3377.}
\label{fig:calibration}
\end{figure}

Fig.~\ref{fig:calibration} presents some typical \sauron\ exposures
taken during commissioning. Panel~\emph{a} shows a small part of an
exposure with the grism taken out, yielding an image of the
micropupils. Panels~\emph{b} and \emph{c} show spectra obtained with
the internal tungsten and neon lamps, respectively.  The former shows
the accurately aligned continuum spectra, while the latter displays
the emission lines used for wavelength calibration. Panel~\emph{d}
presents spectra of the central region of NGC~3377~(see
\S\ref{sec:example}).

\begin{figure}
\epsfxsize=\columnwidth
\fig{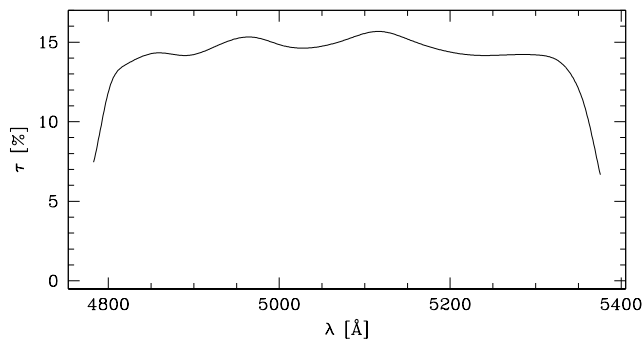}
\caption{Total measured throughput of \sauron\ (including telescope,
atmosphere and CCD transmission).}
\label{fig:throughput}
\end{figure}

We measured flexure by means of calibration exposures taken at 1800~s
interval while tracking, and found $0.2 \pm 0.1$~pixels of shift
between zenith and 1.4 airmasses. We also measured the total
throughput of \sauron, including atmosphere, telescope and detector,
using photometric standard stars observed on a number of photometric
nights. The throughput response is flat at 14.7\%
(Fig.~\ref{fig:throughput}), except for a small oscillation caused by
the intrinsic response of the filter.  This is within a few percent of
the value predicted by the optical design.

\begin{figure}
\epsfxsize=\columnwidth
\fig{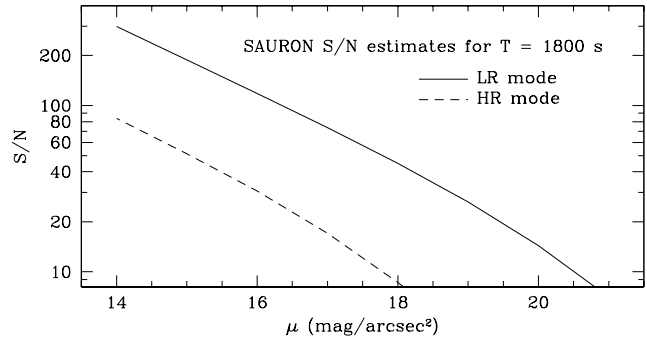}
\caption{Signal-to-noise ratio S/N (per pixel) of individual \sauron\ 
  spectra as a function of surface brightness, for an integration time
  of 1800~s.  The solid line indicates the LR mode and the dashed line
  is for HR mode.}
\label{fig:sn}
\end{figure}

With this efficiency, a signal-to-noise S/N of about 25 per spatial
and spectral element can be reached in 1800~s for a surface brightness
$\mu_V \sim 19$~mag~arcsec$^{-2}$ (Fig.~\ref{fig:sn}). This
corresponds to the typical surface brightness at radii of $\sim$$15''$
in early-type galaxies. By exposing for two hours and co-adding
adjacent spectra in the lower surface brightness outer regions, we can
achieve S/N $\gta 50$, as required for the derivation of the shape of
the LOSVD (see e.g., van der Marel \& Franx 1993).

\section{Data Analysis}
\label{sec:datared}

A typical \sauron\ exposure displays a complex pattern of 1577
individual spectra which are densely packed on the detector (see
Fig.~\ref{fig:calibration}). This pattern defines a three-dimensional
data-set $(\alpha,\delta,\lambda)$ (which we refer to as a
\emph{data-cube}), where $\alpha$ and $\delta$ are the sky coordinates
of a given lenslet, and $\lambda$ is the wavelength coordinate.  We
developed a specific data reduction package for \sauron, with the aim
of (i) producing a wavelength- and flux-calibrated data-cube from a
set of raw CCD exposures, and (ii) extracting the required scientific
quantities (e.g., velocity fields of stars and gas, line-strength
maps, etc.) from the data-cubes, as well as providing tools to display
them.

The software is based on the \texttt{XOasis} data reduction package
(Bacon \etal\ 2000), which includes a library, a set of routines and a
GUI interface. All core programs are written in C ($285\,000$ lines),
while the GUI uses the Tcl/Tk toolkit ($34\,000$ lines). While some
programs are common to both \oasis\ and \sauron, some
instrument-dependent programs were developed specifically for \sauron,
including those for the extraction of spectra (see also Copin
2000). The package is called \texttt{XSauron}, and is described below.

\subsection{CCD preprocessing}
\label{sec:ccd}

The CCD preprocessing is standard, and includes overscan, bias and dark
subtraction.

The overscan value of each line of a raw frame is computed using the
average of 40 offset columns. A median filter of five lines is then
applied to remove high frequency features. The result is subtracted
line by line from the raw exposure.  The master bias is then obtained
by taking the median of a set of normalized biases.

Darks do not show any significant pattern after bias subtraction, thus
a constant value can be removed from each frame. This equals the
nominal value of EEV~12, 0.7~e$^{-1}$~pixel$^{-1}$~hour$^{-1}$. The
dark current therefore plays a negligible role for our typical
integration times of 1800~s.

The \sauron\ software formally includes the possibility to correct for
pixel-to-pixel flat-field variations. As for the EEV~12, apart from a
few cosmetic defects, the pixel-to-pixel variations observed from a
flat-field continuum exposure are quite small (a few percent at most)
and difficult to quantify in our case: it would require measuring the
relative transmission of each pixel on the detector with very high
signal to noise, but, as illustrated in Fig.~\ref{fig:calibration},
the detector is strongly non-uniformly illuminated, especially in the
cross-dispersion direction. We therefore decided not to correct for
pixel-to-pixel flat-field variations. As we will see below, the
resulting error is smoothed out at the extraction stage, where we
average adjacent columns.

The EEV~12 has two adjacent bad columns, one saturated and the other
blind. We interpolate linearly over them. This simple interpolation
scheme is not able to recover the correct intensity value of the
spectra which straddle the bad columns.  However, since the extraction
process takes into account neighbouring spectra (see
\S\ref{sec:extraction}), it is necessary to remove saturated or zero
pixel values to avoid contaminating adjacent spectra.  The
corresponding lenslet is then simply removed at a later stage (see
\S\ref{sec:merge}).

\subsection{Extraction of spectra}
\label{sec:extraction}

The extraction of the individual spectra from the raw CCD frame is the
most instrument-specific part of the data reduction process.  It is
also the most difficult, because of the dense packing of the spectra
on the detector (see Fig.~\ref{fig:calibration}).
Fig.~\ref{fig:crossdisp} illustrates that a simple average of a few
columns around the spectrum peak position will fail to produce
satisfactory results, since it would include light from the wings of
neighbouring spectra. The fact that neighbouring spectra on the
detector are also neighbours on the sky does not help due to the shift
in wavelength (see right panel of Fig.~\ref{fig:maskfit}). We must
therefore accurately remove the contribution of neighbours when
computing the spectral flux.

The method we developed is based on the experience gained with \tiger\
and \oasis, but with significant improvements to reach the accuracy
required here. The principle is to fit an instrumental chromatic model
to high signal-to-noise calibration frames. The resulting parameters
of the fit are saved in a table called the \emph{extraction mask},
which is then applied to every preprocessed frame to obtain the
corresponding uncalibrated data-cube.

\subsubsection{Building the extraction mask}

Building the extraction mask requires three steps: (i) lenslet
location measurement, (ii) cross-dispersion profile analysis and (iii)
global fitting of the instrumental model.

A micropupil exposure (see Fig.~\ref{fig:calibration}\emph{a}) is
obtained with the tungsten lamp on and the grism removed.  Except for
the magnification and some geometrical distortion, this exposure is a
direct image of the lenslet array, and is suitable for deriving its
geometrical characteristics. Each small spot also provides a good
measure of the overall instrumental PSF, i.e., the telescope geometric
pupil convolved by the PSF of the spectrograph with the grism
removed. There are other means to obtain this PSF, e.g., using the
calibration frames, but the micropupils are easier to use since they
are free of any contamination from neighbours, being separated by
large distances (54~pixels, cf.\ the 4.8~pixel separation between
adjacent spectra).

After automatic thresholding, the centroids of the micropupils are
computed and the basic parameters of the lens array (lenslet positions
and tilt with respect to the detector columns) are derived.  The
geometrical distortion of the spectrograph can be evaluated using the
observed difference between the measured micropupil position and the
theoretical one as a function of radius. In contrast to the case of
\oasis, the geometrical distortion of \sauron\ is negligible over the
detector area.

We now describe in more detail the model of the cross-dispersion
profile and its variation over the field of view.  For a good
treatment of the pollution by neighbouring spectra (and optimal
extraction), we are only interested in the profile of the PSF along
the cross-dispersion direction, while the dispersion profile directly
sets the spectral resolution of the observations. In order to be
compatible with the observed cross-dispersion profile of a continuum
spectrum, the PSF must be collapsed along the dispersion direction to
obtain its \emph{integrated cross-dispersion profile} (ICDP).  For
convenience, we make no formal distinction between the PSF and its
ICDP.

\begin{figure}
\epsfxsize=\columnwidth
\fig{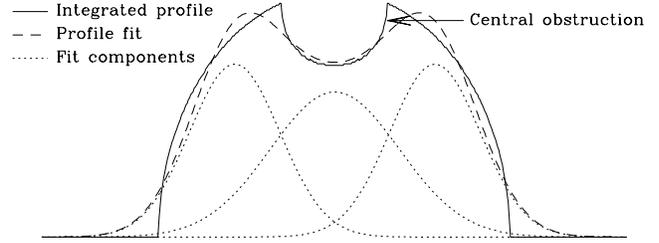}
\caption{Integrated cross-dispersion profile ${\cal G}$ of the geometrical
  micropupil (solid line) and its fit (dashed line) with
  three Gaussian functions (dotted lines).}
\label{fig:geopup}
\end{figure}

The overall ICDP ${\cal P}$ is the convolution of the geometrical
micropupil profile ${\cal G}$ with the PSF of the spectrograph ${\cal
F}$.  The former can be computed easily knowing the primary mirror and
central obstruction size, the aperture of the lenslet on the sky, and
the aperture of the camera.  Accordingly, the micropupil geometrical
size is 2.6 and 0.9~pixel for the LR and HR modes, respectively, and
its ICDP---uniform over the CCD area---is shown in
Fig.~\ref{fig:geopup}. For ease of computation, we represent the
profile ${\cal G}$ by the sum of three Gaussian functions.

${\cal F}$, the second contribution to ${\cal P}$, is caused by
imperfections in the lenslet array and spectrograph, and is likely to
change with the instrument focus and setup, and to vary from one
lenslet to the other.  While the geometrical pupil is constant over
the CCD area, the observed FWHM of the ICDP increases with distance
from the optical center, and from red to blue on a given spectrum. We
thus decided to model the spectrograph PSF ${\cal F}_{i}$ for lenslet
$i$ as the convolution of a fixed part ${\cal F}^{\star}$ (the kernel)
with a chromatic local component ${\cal F}_{i}(\lambda)$. Overall, the
cross-dispersion profile ${\cal P}_{i}(\lambda)$ of spectrum $i$ at
wavelength $\lambda$ is approximated by:
\begin{equation}
  \label{eq:1}
  {\cal P} = {\cal G} \otimes {\cal F}^{\star} \otimes {\cal F}_{i}(\lambda).
\end{equation}
Due to the absence of a dispersing element, ${\cal F}_{i}(\lambda)$
cannot be retrieved from the micropupils, but if we restrict our
analysis to the micropupils very close to the optical center, where
${\cal F}_{i}(\lambda)$ is Dirac-like, ${\cal G} \otimes {\cal
F}^{\star}$ can be derived. A sum of three centered Gaussians gives a
precise approximation of the kernel.  The corresponding fit, which
takes into account the pixel integration, is performed simultaneously
on the ten central micropupils. This procedure results in an accurate
kernel shape without being affected by undersampling.

\begin{figure}
\epsfxsize=\columnwidth
\fig{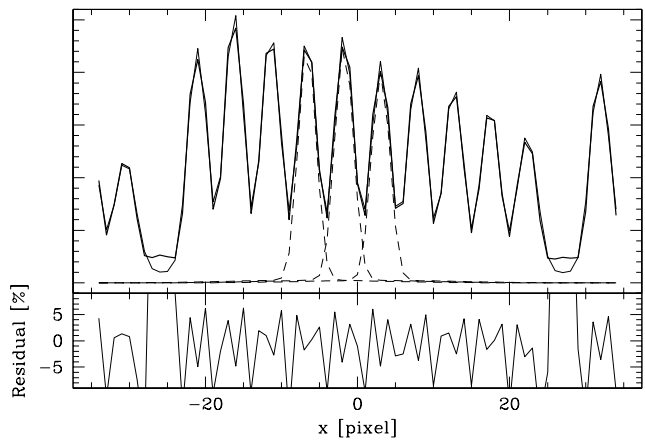}
\caption{Upper panel: example of a cross dispersion profile derived
  from a continuum exposure (thick solid line) and its fit by the
  model described in text (thin solid line).  To illustrate the level
  of overlap, only the central three fitted profiles are shown (dashed
  lines). Lower panel: relative residual of the fit (in~\%). }
\label{fig:crossdisp}
\end{figure}

In the second step, the last component ${\cal F}_{i}(\lambda)$ of
Eq.~(\ref{eq:1}) must be derived from a spectroscopic exposure, using
the previously estimated ${\cal G} \otimes {\cal F}^{\star}$ kernel.
For this purpose, we use a high signal-to-noise continuum exposure
obtained with the tungsten lamp (see Fig.~\ref{fig:calibration}{\emph
b}). A single Gaussian is enough to describe this component, but the
dense packing requires the simultaneous fit of a group of
cross-dispersion profiles, and makes the process CPU intensive.  The
fit is accurate to a few percent, as shown in
Fig.~\ref{fig:crossdisp}.  Given the slow variation of $\sigma_{{\cal
F}_{i}}(\lambda)$ with $\lambda$, we perform this fit only once
for each ten CCD lines.  At each analysed line $y$, the position $x$
and width $\sigma$ of each peak is saved in a file for later use as
input data for the global mask fitting.

The above construction of the ICDP produces a large collection of
$(x,y,\sigma)$ giving position and characteristic width of all the
cross-dispersion peaks over the entire detector. The final step is to
link these unconnected data to individual spectra.  By contrast to the
two previous steps, which involve only local fitting, this last
operation requires the fit of a global instrumental chromatic model,
based on an \emph{a priori} knowledge of basic optical parameters
measured in the laboratory: the camera and collimator focal lengths,
and the number of grooves of the grism. It also requires the fitting
of extra parameters which depend on the setup of the instrument
and/or were not previously measured with enough accuracy: (i) the
three Euler angles of the grism, (ii) the chromatic dependence of the
collimator and camera focal lengths, (iii) the distortion parameters
of the collimator and camera optics as well as their chromatic
dependence, and (iv) the distortion center of the optical system.

\begin{figure}
\epsfxsize=\columnwidth
\fig{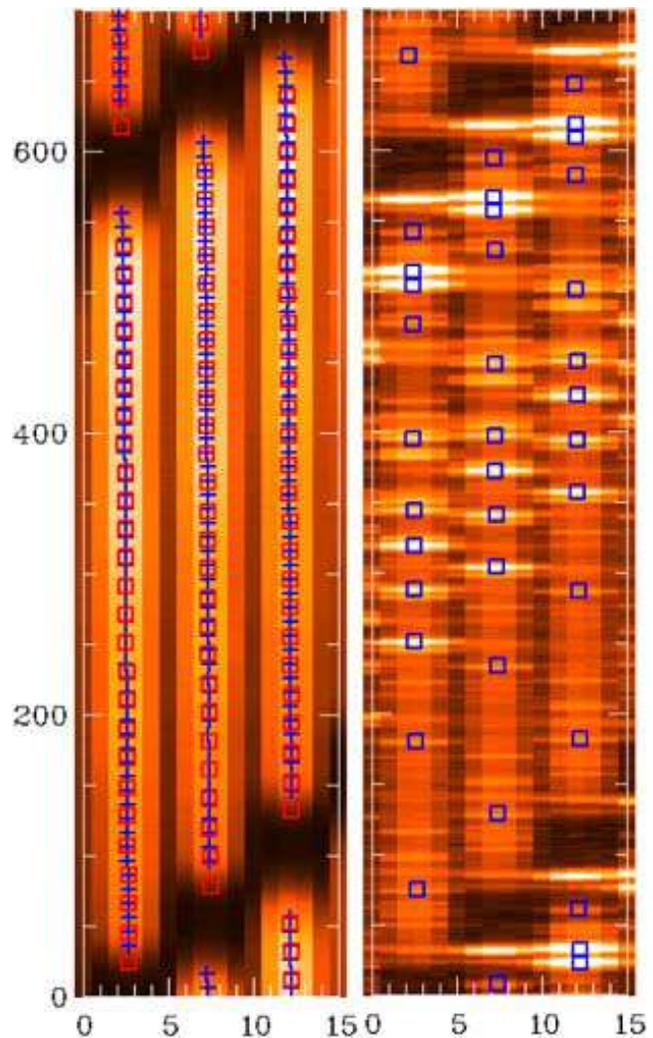}
\caption{Example of mask-fitting results. Left panel: a small part of
  a continuum exposure with the derived locations of the spectra
  (crosses) and the corresponding fitted values (squares).  Right
  panel: the corresponding arc (Neon) exposure and fitted emission
  lines (squares). The images have been expanded along the
  cross-dispersion axis for clarity.  }
\label{fig:maskfit}
\end{figure}

The fitting of the instrumental model requires the $(x,y,\sigma)$
measurements of the cross-dispersion profiles and an arc exposure with
its corresponding emission-line wavelengths tabulated. The latter is
needed to constrain efficiently the wavelength-dependent quantities.
The fit uses a simple ray-tracing scheme, with `ideal' optical
components, to predict the $(x,y)$ location of each spectrum as
function of wavelength $\lambda$.  The predicted locations of the
continuum spectra and the arc emission lines are compared with the
corresponding data and differences are minimized by the fit
(Fig.~\ref{fig:maskfit}).  Convergence is obtained after a few hundred
iterations with a simplex scheme. The model is then accurate to $\lta
0.1$~pixel RMS.

The ICDP has a noticeable chromatic variation, causing a $\sim$$15$\%
relative increase of its width from the red (5400~\AA) to the blue end
(4760~\AA).  The slope of this chromatic variation is fit with a
linear relation for each individual spectrum.  The last improvement is
to allow for tiny offsets and rotations of the spectra which are not
included in our simple mathematical model of the instrument, in order
to get the most precise mask position.

The final mask describes the best model of the instrument, including
second-order lens-to-lens adjustments. This mask is accurate for a
full run, and has been computed precisely for the first run. 
The setup can vary slightly from run to run, but most of the
optical parameters (e.g., the focal length) are stable.  We therefore
adopt a simplified procedure using the mask from the first run as
starting point, and fitting only the parameters that are likely to
change with the setup, such as the global translation and rotation of
the detector.

\subsubsection{Extracting spectra}

The preparatory work described in the above has only one objective: to
extract accurate spectro-photometric spectra from the preprocessed CCD
frames.  The procedure requires (i) the lenslet position in sky
coordinates, (ii) the location of the spectra in CCD coordinates and
the corresponding wavelength coordinates, and (iii) the ICDP of the
spectra.  All this information is stored in the instrumental model
contained in the extraction mask. However, before the extraction
process itself, we must take into account possible small displacements
of the positions of the spectra relative to the extraction mask. These
offsets are due to a small amount of mechanical flexure between the
mask exposure---generally obtained at zenith---and the object
exposure.  They are estimated using the cross-correlation function
between arc exposures obtained before and/or after the object exposure
and the arc exposure used for the mask creation.\looseness=-2

For each lenslet $i$ referenced in the extraction mask, the
instrumental model gives the $(x_i,y_i,\lambda(y_i))$ geometrical
position on the CCD of the corresponding spectrum. At each position,
one must compute the total flux which belongs to the associated
spectrum (see Fig.~\ref{fig:crossdisp}).  This step is critical, as
any error will affect the spectro-photometric accuracy of the
extracted spectrum. We developed an algorithm to remove the pollution
induced by the two neighbouring spectra and to compute the flux in an
optimal way (as defined by Horne 1986 and Robertson 1986).  The
algorithm uses the ICDP of the three adjacent peaks (the current
spectrum and its two neighbours). The three profiles are normalized to
the intensity of the peaks, and the estimated cross-dispersion profile
of the neighbours is subtracted from the main peak in a dual-pass
procedure. Finally, the total flux of the main peak is calculated in
an extraction window of five pixels width, using a weighted sum of the
intensity of the pixels, corrected for the pollution by neighbours.
The total variance of the integrated flux is saved as an estimate of the
spectrum variance.

\begin{figure}
\epsfxsize=\columnwidth
\fig{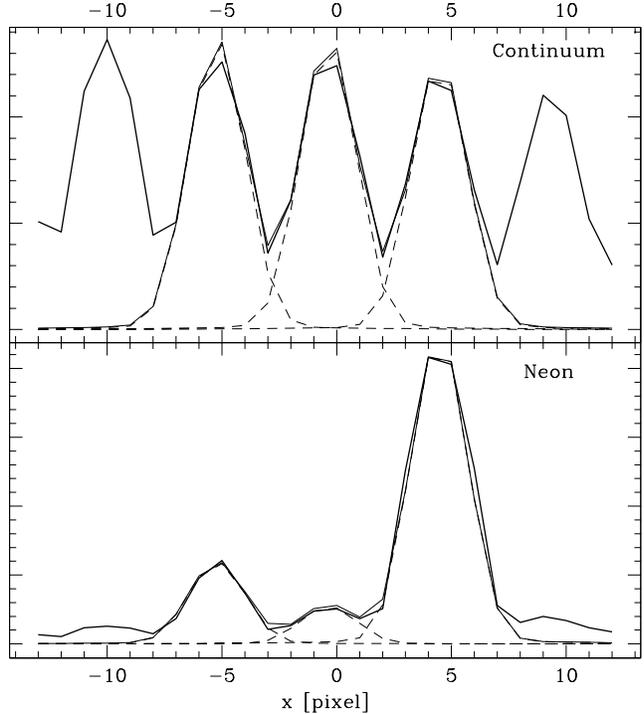}
\caption{Examples of flux computation in the extraction process on
  continuum (upper panel) and neon arc (lower panel) exposures. Thick
  solid line: data. Thin solid line: sum of the estimated
  central-plus-neighbours cross-dispersion profiles. Dashed lines:
  individual cross-dispersion profiles.}
\label{fig:extract}
\end{figure}

Examples of this process are shown in Fig.~\ref{fig:extract}, which is
repeated along each pixel of the current spectrum. The result is saved
in a specific binary structure called the \tiger\ data-cube. This contains
all the extracted intensity spectra and their corresponding variance
spectra. Each pair of object \& noise spectra is referenced by a lens
number. The data-cube is attached to a FITS table containing the
corresponding lens coordinates and any other pertinent data.

\subsection{Wavelength calibration}

The wavelength calibration is standard, and uses neon arc exposures
which contain 11 useful emission lines in the \sauron\ wavelength
range.  Since the extraction process is based on a chromatic model,
the raw extracted spectra are already pre-calibrated in wavelength
(scale $\lambda'$). Therefore, the final wavelength calibration only
needs to correct for second order effects which were not included in
the instrumental model, and is thus straightforward and robust.  A
cubic polynomial is fitted to the $(\lambda'_k,\lambda_k)$ points, and
is saved for rebinning of the spectra. The RMS residual after
rebinning is $\lta 0.1$~\AA, which corresponds to 0.1~pixel or $\sim
5$~\kms.

\subsection{Flat-fielding}

The flat-field is a complicated function of spatial coordinates
$(x,y)$ and wavelength $\lambda$, which depends on the lenslet,
filter, optics and detector properties. A continuum exposure obtained
with the internal tungsten lamp gives featureless spectra, which can
be used to remove the wavelength signature of the flat-field. However,
the internal calibration illumination is not spatially uniform and
cannot be used to correct for the spatial variation of the
flat-field. For this, we use a twilight exposure.  The final
flat-field data-cube is the product of the spectral flat-field
(obtained by dividing a reference continuum by the continuum
data-cube) and the spatial flat-field variations measured on the
twilight data-cube.\looseness=-2

The flat-field correction displays a large-scale gradient from the
upper to the lower regions of the detector, with a maximum of 20\%
variation over the whole field of view. At both edges of the filter,
the very strong gradient of transmission is difficult to correct for,
so we truncate each spectrum at the wavelength where the derivative of
the transmission increases above a chosen threshold.

\subsection{Cosmic ray removal}

In imaging and long-slit spectroscopy, cosmic rays are generally
removed directly from the CCD frame.  The standard procedures are much
more difficult to apply in the case of \tiger-type integral-field
spectroscopy, since the high contrast arising from the organization of
spectra on the detector prevents the use of standard algorithms.
However, we can exploit the three-dimensional nature of the data-cube
$(x, y, \lambda)$ to our advantage: using a continuity criterion in
\emph{both} the spatial and the spectral dimension on a
wavelength-calibrated data-cube, we can detect any event which is
statistically improbable.  In practice, each spectrum is subtracted
from the median of itself and its eight nearest spatial neighbours.
The resulting spectrum, which corresponds to spatially unresolved
structures, is filtered along the wavelength direction with a
high-pass median filter.  Statistically deviant pixels are then
flagged as `cosmic rays' and replaced by the values in the spatially
medianed spectrum. The affected pixels are flagged also on the
corresponding noise spectrum.

\subsection{Uniform spectral resolution}

At this stage, the spectra may have slightly different spectral
resolutions over the field of view, due to, e.g., optical distorsions
in the outer regions, CCD-chip flatness defects, non-orthogonality
between CCD-chip mean plane and optical axis.  In the \sauron\ LR
mode, the instrumental spectral dispersion $\sigma_{\mathrm{inst}}$
typically increases from $\sim 90$~\kms\ close to the center to $\sim
110$~\kms\ at the edge of the field. We need to correct for this
variation before we can combine multiple exposures of the same object
obtained at different locations on the detector.  For each
instrumental setup, we derive the local instrumental spectral
resolution over the field of view from the auto-correlation peak of a
twilight spectrum.  The spectral resolution of the data-cubes are then
smoothed to the lower resolution using a simple Gaussian convolution.

\subsection{Sky subtraction}

Sky subtraction is straightforward: we simply compute the median of
the 146 dedicated sky spectra and subtract it from the 1431 object
spectra.  The accuracy of sky subtraction, measured from blank sky
exposures, is 5\%. We also checked that the sky emission-line doublet
[NI] $\lambda$5200~\AA\ is removed properly.\looseness=-2

\subsection{Flux calibration}

Flux calibration is classically achieved using spectro-photometric
standard stars observed during the run as references. These stars are
often dwarfs whose spectra generally exhibit few absorption features,
although most of them show strong Balmer lines. The available
tabulated flux curves usually have poor spectral resolution (sampling
at a few tens of \AA). Furthermore, systematic wavelength calibration
errors of up to 6~\AA\ at 5000~\AA\ are not uncommon. This prevents an
accurate flux calibration of the \sauron\ spectra which cover only a
short wavelength region.

We therefore decided to favor flux-calibrated Lick (K0III) standards,
previously observed at high spectral resolution, as \sauron\
flux-standard stars. These high-resolution stellar spectra are first
convolved to the \sauron\ resolution and corrected to zero
(heliocentric) velocity shift (using the Solar spectrum as a
reference). The spectra are then used in place of the classical
standard flux tables. This procedure provides an accuracy of about
5\%.

Alternatively the relative transmission curve can also be measured
using special stars with featureless spectra in the \sauron\
wavelength range. The derived curve is then scaled to absolute
flux using photometric standard stars.

\subsection{Merging multiple exposures}
\label{sec:merge}

The total integration time on a galaxy is usually split into 1800~s
segments. This value was chosen to limit the number of cosmic ray
events and to avoid degradation of the spectral resolution caused by
instrumental flexure (see \S\ref{sec:commissioning}).  Another reason
to split the integration in at least two exposures is to introduce a
small spatial offset between successive exposures. These offsets are
typically a few arcsec and correspond to a few spatial sampling
elements. The fact that the same region of the galaxy is imaged on
different lenslets is used in the merging process to minimise any
remaining systematic errors, such as flat-field residuals, and to deal
with a column of slightly lower quality lenslets.  Furthermore, when a galaxy
is larger than the \sauron\ field of view, it is also necessary to
mosaic exposures of different fields, and the merging process has to build
the global data-cube for the larger field from the individual data-cubes.

The merging process consists of four successive steps: (i) truncation
of the spectra to a common wavelength range, (ii) relative centering
of all individual data-cubes, (iii) computation of normalisation and
weight factors, and (iv) merging and mosaicing of all data-cubes.

The spectral range varies over the field due to the tilt of the filter
(see \S\ref{sec:instrument}). The spectral range common to all spectra
in the \sauron\ field is only 4825--5275~\AA\ 
(Fig.~\ref{fig:throughput}). In some cases, a larger spectral coverage
is required, which we can reach by ignoring some lenses, i.e., by
shrinking the field.  For instance, a wavelength range of
4825--5320~\AA\ implies a reduction of the field of view by 36\%.

After truncating all spectra to a common wavelength domain, we build
integrated images for each individual data-cube by summing the entire
flux in each spectrum and interpolating the resulting values. These
\emph{reconstructed images} are then centered using a
cross-correlation scheme, taking care that the result is not affected
by differences in seeing.

In principle, one should take into account the effects of
differential atmospheric refraction before reconstructing images. By
contrast to long-slit spectroscopy, here it is possible to correct for
this effect if the integration time is short enough. The method is
described in Arribas et al.\ (1999), and was applied successfully to
\tiger\ and \oasis\ data (e.g., Emsellem at al.\ 1996). The galaxies
in our representative sample will be observed at an airmass less than
1.3 (Paper II), so that the differential atmospheric refraction is
sufficiently small (0\farcs16) to be neglected.

To optimize the use of telescope time, we do not observe photometric
standards after each galaxy exposure. Changing observing conditions
may cause normalisation errors in the flux calibration. We therefore
renormalise all individual data-cubes using the common spatial area.
During the merging process, each data-cube is also weighted to
maximise the signal-to-noise ratio by an optimal summation
scheme.\looseness=-2

Finally, each data-cube is interpolated onto a common square grid.
Since the different exposures are dithered by small spatial offsets,
we use a finer grid to improve the spatial sampling of the combined
data-cube (`drizzling' technique).  At each grid point, the spectrum
of the resulting merged data-cube is simply the weighted mean of all
spectra available at this location.  The variance of the merged
spectra are derived from the variances of the individual
spectra.\looseness=-2

\subsection{Palantir pipeline}

The representative sample of galaxies that we are observing with
\sauron\ (see Paper~II) will generate an estimated total of 40~Gb of
raw CCD frames and 20~Gb of reduced images and data-cubes.  Some of
the algorithms presented in the previous sections will evolve in time,
to take into account the improved understanding of the instrument that
comes with repeated use.  In order to ease the re-processing of
data-sets, we have developed a special pipeline environment, called
\texttt{Palantir}.  The main components of this are the \sauron\ 
software described above, a MySQL database, a Tcl/Tk interface for
querying the database and sending data to the pipeline, and the OPUS
pipeline management system (Rose \etal\ 1995).  The database contains
header information for the raw data, processing parameters and the log
of data reduction.  Using the interface, the desired class of data can
be quickly selected and processed with the correct calibration files.
OPUS was developed at STScI for managing the HST pipeline and provides
tools for monitoring data in a pipeline.  The extraction procedure is
time consuming, so it is important to maximize the efficiency and the
automation: OPUS does this by allowing parallel processing, on
multiple CPUs if necessary.

The output of \texttt{Palantir} is one data-cube for each observed
object with the instrumental signature removed. Each data-cube
includes also a noise variance estimate. These `final' data-cubes are
the starting point for the scientific measurements.  Below we
summarize the main procedures we employ; more details will be found in
subsequent papers that will address specific scientific questions, as
the procedures sometimes need to be tailored to be able to deal with
particular objects (e.g., counter-rotating disks).

\section{Data analysis software}
\label{sec:dataana}

For scientific analysis, we need to measure a few key parameters from
each spectrum of the final data-cube.  These numbers, which could be,
e.g., a flux, a velocity, or a line ratio, are attached to the spatial
location of the corresponding spectrum. We then build two-dimensional
images of these quantities, giving, e.g., velocity fields,
line-strength maps, using a bilinear (or bicubic) interpolation
scheme.  The \texttt{XSauron} software allows us to display these
images and examine the spectrum at a certain location interactively.

\subsection{Photometry}
\label{sec:photo}

A simple integration over the full wavelength range of the flux
calibrated \sauron\ datacubes gives the reconstructed image of the
target. This image gives the reference location of each spectrum.  The
kinematic and line-strength maps are by definition perfectly matched
to this continuum intensity map.  The reconstructed image can be used
also to perform a photometric analysis of the object. In most cases
this is not very useful since HST and/or ground-based broad-band
images with finer sampling and a larger field of view will be
available. However, a comparison of a high-resolution direct image
with the \sauron\ reconstructed image provides an excellent test of
the scientific quality of the instrument and the reduction algorithms.
Any errors in the end-to-end process such as flat-field inaccuracies,
or data extraction errors, will affect the reconstructed image and
will be apparent in the comparison.

\begin{figure}
\epsfxsize=\columnwidth
\fig{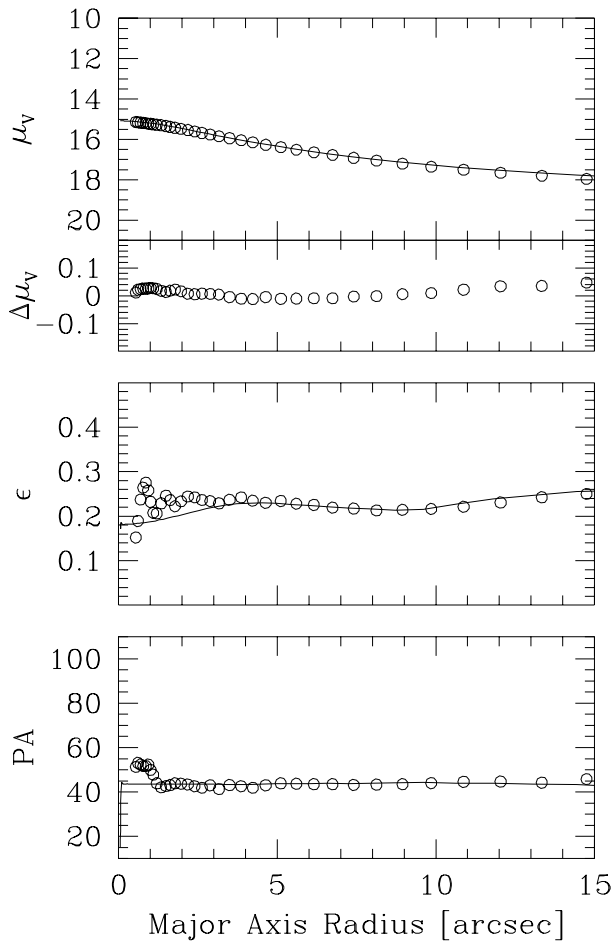}
\caption{A comparison of two photometric analyses of NGC~4365.  The
first (solid line) was obtained using ellipse fitting on HST/\wfpc2\
F555W data convolved with a 2\farcs1 PSF.  The second (open circles)
was derived in the same way from the \sauron\ reconstructed flux image
in the 4850--5230 \AA\ wavelength range. Top to bottom show profiles
of surface brightness in magnitudes ($\mu_v$), $\Delta\mu = \mu_{\rm
SAURON}-\mu_{\rm HST}$, ellipticity ($\epsilon$) and position angle of
the major axis (PA). }
\label{fig:comp4365}
\end{figure}

We have carried out a photometric comparison for the E3 galaxy
NGC~4365, observed with \sauron\ in March 2000 (Davies et al.\
2001). We used the GALPHOT package (Franx, Illingworth \& Heckman
1989) to fit ellipses to the reconstructed image, and derive the
standard photometric profiles.  The same process was repeated on a
high resolution image obtained from the HST/\wfpc2\ archive (\#5454 PI
Franx and \#5920 PI Brache). We used the F555W filter since it is
close to the \sauron\ bandpass. The HST image was convolved with a
Gaussian of 2\farcs1 which is a rough estimate of the \sauron\ spatial
resolution. The comparison is shown in Fig.~\ref{fig:comp4365}.  The
\sauron\ data was normalized to the total flux of the HST image. The
profiles agree very well: the RMS difference, excluding the central
$2''$ (to avoid including any residual difference in spatial
resolution) is only 0.016 in magnitude, 0.012 in ellipticity, and
1.0$^\circ$ in position angle of the major axis.

\subsection{Stellar kinematics}
\label{sec:fcq}

Stellar kinematical quantities are derived with a version of the
Fourier Correlation Quotient (FCQ) algorithm of Bender (1990),
implemented in the \texttt{XSauron} software. Other algorithms,
including Fourier Fitting (van der Marel \& Franx 1993) and Unresolved
Gaussian Decomposition (Kuijken \& Merrifield 1993) will be
incorporated in the near future.

During an observing run, we obtain a set of stellar exposures with
\sauron, e.g., G, K and M giants, which are fully reduced providing a
corresponding set of stellar data-cubes. For each data-cube, we
measure FWHM$_{\star}$, the full-width-at-half-maximum of the PSF of
the reconstructed stellar image. All spectra within 1~FWHM$_{\star}$
of the brightest spectrum in the data-cube are summed, resulting in a
single high-S/N stellar spectrum.  All spectra---galaxy and stellar
templates---are then rebinned in $\ln\lambda$, to have a scale
proportional to velocity. A data-cube containing the LOSVDs of the
galaxy is then derived via FCQ using a single stellar template,
typically a K0 giant).  We then measure the mean velocity $V$ and
velocity dispersion $\sigma$ from the LOSVDs, and these values are
used to build a linear combination of all observed stellar templates,
the so-called `optimal stellar template', which matches best the
galaxy spectra. This optimal template is derived by a linear fit in
real $\lambda$-space, and include a low-order polynomial continuum
correction when needed.  The LOSVDs are then rederived with the
optimal template. We finally reconstruct maps of $V$ and $\sigma$ as
well as the Gauss-Hermite moments $h_3$ and $h_4$ (see van der Marel
\& Franx 1993; Gerhard 1993) by measuring these values on the final
LOSVD data-cube.

\subsection{Gas kinematics}

The construction of an optimal template, as described in
\S\ref{sec:fcq}, can be done individually for each spectrum in a
galaxy data-cube, resulting in a corresponding `optimal template'
data-cube. If the presence of emission lines is suspected,
contaminated wavelength regions can be masked easily during the
template-fitting process in $\lambda$-space. The broadened optimal
template spectra are then subtracted from the galaxy spectra yielding
a data-cube (presumably) free of any contribution from the stellar
component. Emission lines, if present, are then easily measurable, and
their fluxes, velocities and FWHM can be derived, assuming, e.g., a
Gaussian shape for each line. We then build the corresponding maps of
the gaseous distribution, kinematics, as well as line ratios (e.g.,
H$\beta$/[O\textsc{iii}]). In some cases, emission lines will be
strong enough to perturb the derivation of the stellar LOSVDs. When
this happens, we subtract the fitted emission lines from the original
galaxy spectra, and then recompute (a third time, see \S\ref{sec:fcq})
the LOSVDs and their corresponding velocity moments.

\subsection{Line-strengths}

Line-strength indices are measured in the Lick/IDS system (Worthey
1994). The wavelength range of the current \sauron\ setup allows
measurements of the H$\beta$, Fe~5015, Mg$b$ and Fe~5270 indices. In
order to transform our system of line-strengths onto the standard Lick
system, we broaden each flux calibrated spectrum in the final
data-cube to the Lick resolution of 8.4~\AA\/ (FWHM) (Worthey \&
Ottaviani 1997) and then measure the indices. All index measurements
are then corrected for internal velocity broadening of the galaxies
(e.g., Kuntschner 2000) using an `optimal stellar template' and the
velocity dispersion derived from each individual spectrum (see
\S~5.1).\looseness=-2

Due to differences in the continuum shape between our data and the
original Lick setup, there remain generally some small systematic
offsets for some indices. In order to monitor these, we
observe a number ($>10$) of Lick standard stars in each run,
and compare our measurements with the Lick data (Trager \etal\
1998). This ensures inter-run consistency and also allows us to remove
the systematic offsets.

\begin{figure*}
\epsfxsize=16.8cm
\fig{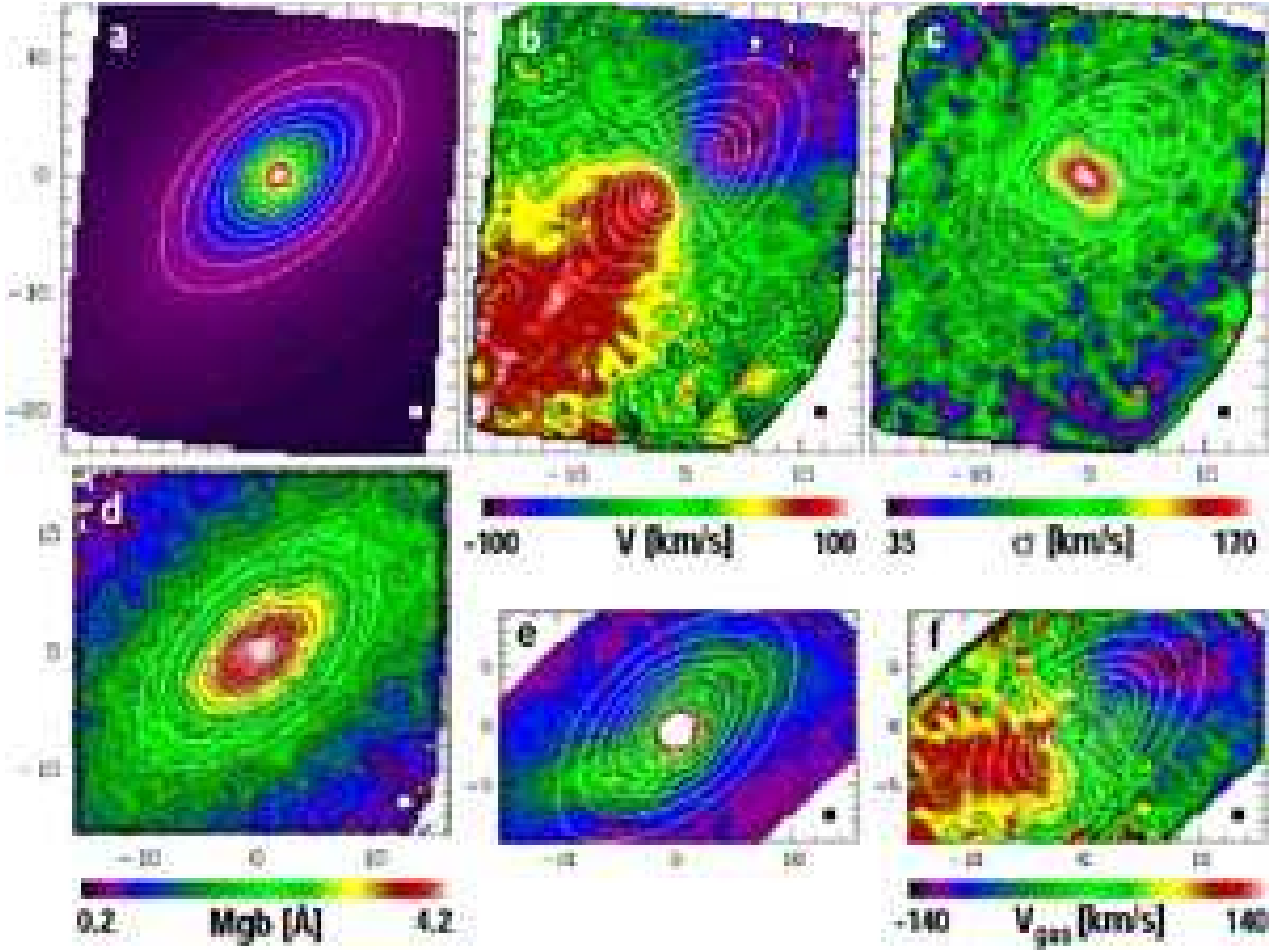}
\caption{\sauron\ observations of the E6 galaxy NGC~3377. a)
  Reconstructed total intensity.  b) Mean velocity
  $V$. c) Velocity dispersion $\sigma$.
  d) Stellar Mg$b$ index.  e) Gaseous
  total intensity ([O\textsc{iii}] $\lambda$5007).  f) Gas velocity field.
  Isophotes of the reconstructed image are superimposed on each frame.  
  The small square at the lower right corner of each frame gives the size
  of the lenslets ($0\farcs94\times0\farcs94$), 
  and the field shown is $30''\times39''$.
  These maps are based on a $4\times 1800$~s exposure. 
  No spatial binning has been applied at this stage. The stars show a
  striking rotating disk pattern with the spin axis misaligned about
  $10^\circ$ from the photometric minor axis, indicating the galaxy is
  triaxial.  The gas also reveals strong non-axisymmetric structures
  and motions.  The Mg$b$ isophotes follow the continuum light.}
\label{fig:n3377}
\end{figure*}

\section{SAURON observations of NGC3377}
\label{sec:example}

The companion paper (Paper~II) describes the first results of a
systematic survey of nearby early-type galaxies with \sauron. Here, we
illustrate the power and versatility of \sauron\ by presenting
observations of the galaxy NGC~3377. This E6 galaxy in the Leo~I group
has a steep central luminosity profile (Faber \etal\ 1997), a total
absolute magnitude $M_B=-19.24$, and has been probed for the kinematic
signature of a massive central black hole from the ground (Kormendy
\etal\ 1998).

We observed NGC~3377 with \sauron\ on February 17 1999, under
photometric conditions, at a median seeing of 1\farcs5 (see Copin
2000). We exposed for a total of $4\times 1800$~s.
Fig.~\ref{fig:n3377} shows the resulting maps. The velocity field
shows a regular pattern of rotation, but the rotation axis is
\emph{misaligned} by about $10^\circ$ from the photometric minor axis.
This is the signature of a triaxial intrinsic shape. 
The Mbg isocontours seem to follow the continuum isophotes, but not
the contours of constant dispersion.
The ionised gas distribution and kinematics show strong departures
from axisymmetry, this component exhibiting a spiral like morphology.
This could be another manifestation of the triaxiality already
probed via the stellar kinematics, and may be the signature of a
bar-like system. This would not be too surprising since we expect
anyway about two third of S0 galaxies to be (weakly or strongly)
barred. This suggest that early-type systems with steep central
luminosity profiles are not necessary axisymmetric.
In this respect, it will
be interesting to establish the mass of the central black hole.

The full analysis of the \sauron\ data for NGC~3377 will be presented
elsewhere (Copin \etal, in prep); it will include dynamical modeling
which also incorporates \oasis\ observations of the central $2''\times
4''$ with a spatial resolution of 0\farcs6.

\section{Concluding remarks}
\label{sec:conclusions}

The number of integral-field spectrographs operating on 4--8~m class
telescopes is increasing rapidly, and others are under construction or
planned, including \texttt{SINFONI} for the VLT (Mengel
\etal\ 2000) and \texttt{IFMOS} for the NGST (Le F\`evre \etal\
2000). Much emphasis is being put on achieving the highest spatial
resolution, by combining the ground-based spectrographs with adaptive
optics, in order to be able to study e.g., galactic nuclei. \sauron\
was designed from the start to complement these instruments, for the
specific purpose of providing the wide-field internal kinematics and
line-strength distributions of nearby galaxies.

The plan to develop \sauron\ was formulated in the summer of 1995 and
work commenced in late spring 1996. The instrument was completed in
January 1999. This fast schedule was possible because \sauron\ is a
special-purpose instrument with few modes, and because much use could
be made of the expertise developed at Observatoire de Lyon through the
building of \tiger\ and \oasis. While \sauron\ was constructed as a
private instrument, plans are being developed to make it accessible to
a wider community.

\sauron's large field of view with 100\% spatial coverage is unique,
and will remain so for at least a few years. For comparison,
\integral, which also operates on the WHT (Arribas \etal\ 1998), uses
fibers and has a wider spectral coverage than \sauron.  It has a
mode with a $34'' \times 29''$ field of view and $2\farcs7$ sampling,
but the spatial coverage is incomplete and the throughput is
relatively modest. \vimos\ (on the VLT in 2001, see Le F\`evre \etal\
1998) will have an integral-field spectrograph with spectral
resolution relevant for galaxy kinematics and with a $27'' \times
27''$ field sampled with $0\farcs67\times0\farcs67$ pixels. The mode
with a $54''\times54''$ field of view has insufficient spectral
resolution. \sauron's $33''\times 41''$ field, together with its high
throughput, make it the ideal instrument for studying the rich
internal structure of nearby early-type galaxies.

\medskip
\thanks It is a pleasure to thank the ING staff, in particular Rene
Rutten and Tom Gregory, as well as Didier Boudon and Rene Godon for
enthusiastic and competent support on La Palma. The very efficient
\sauron\ optics were designed by Bernard Delabre and the software
development benefitted from advice by Emmanuel and Arlette P\'econtal.
RLD gratefully acknowledges the award of a Research Fellowship from
the Leverhulme Trust.  The \sauron\ project is made possible through
grants 614.13.003 and 781.74.203 from ASTRON/NWO and financial
contributions from the Institut National des Sciences de l'Univers,
the Universit\'e Claude Bernard Lyon I, the universities of Durham and
Leiden, and PPARC grant `Extragalactic Astronomy \& Cosmology at
Durham 1998-2002'.

{}

\end{document}